%% file: sigconf.tex
\documentclass[manuscript,acmsmall,nonacm]{acmart}
\setcopyright{none}
\usepackage{svg}
\usepackage{subcaption}
\usepackage{colortbl,tabularx,array}

\AtBeginDocument{%
  }

\setcopyright{acmlicensed}
\copyrightyear{2018}
\acmYear{2018}
\acmDOI{XXXXXXX.XXXXXXX}
\acmConference[Conference acronym 'XX]{Make sure to enter the correct
  conference title from your rights confirmation email}{June 03--05,
  2018}{Woodstock, NY}
\acmISBN{978-1-4503-XXXX-X/2018/06}

\begin{document}

\title[Making Smart Homes Fit for Autistic Children]{Making Smart Homes Fit for Autistic Children: Family Work and Monitored Independence}

\author{Tingting Xiang}
\email{xiangtt4179@gmail.com}
\affiliation{%
  \institution{Chongqing Jiaotong University}
  \city{Chongqing}
  \state{Chongqing}
  \country{China}
}

\author{Weihan Jiang}
\email{15944601611@163.com}
\affiliation{%
  \institution{Independent Researcher}
  \city{Austin}
  \state{Texas}
  \country{USA}
}

\author{Hrishitva Patel}
\email{hrishitva.patel@utsa.edu}
\affiliation{%
  \institution{University of Texas at San Antonio}
  \city{San Antonio}
  \state{Texas}
  \country{USA}
}

\author{Haiqi Shen}
\email{hs7619206@gmail.com}
\affiliation{%
  \institution{Shanghai AI Laboratory}
  \city{Shanghai}
  \state{Shanghai}
  \country{China}
}

\author{Tianzhi He}
\email{tianzhi.he@utsa.edu}
\affiliation{%
  \institution{University of Texas at San Antonio}
  \city{San Antonio}
  \state{Texas}
  \country{USA}
}

\author{Kyrie Zhou}
\email{kyrie.zhou@utsa.edu}
\affiliation{%
  \institution{University of Texas at San Antonio}
  \city{San Antonio}
  \state{Texas}
  \country{USA}
}

\renewcommand{\shortauthors}{Xiang et al.}

\begin{abstract}
General-purpose smart-home technologies are increasingly present in family life, yet how they support or disrupt autistic children and families with autistic children remains underexamined. We conducted a multi-source qualitative study drawing on five caregiver interviews, Xiaohongshu (also known internationally as RedNote) posts explicitly situated in autism contexts, and contextual posts from Xiaohongshu and Weibo. We find that families create support for autistic children by adapting environmental outputs, prompts, and interaction channels, while sensory intensity, communicative rules, and opaque operating states can turn the same technologies into barriers. Support also depends on continuing family work, including translation, configuration, modification, maintenance, and boundary-setting. Remote monitoring enables conditional independence while generating tensions among safety, privacy, and control. We contribute three concepts, including relational fit, family work as support infrastructure, and monitored independence, for understanding accessible domestic technology. We derive design implications for adjustable interaction, legible and recoverable failure, lower-burden configuration, and negotiable monitoring.
\end{abstract}



\maketitle

\input{introduction}

\input{related-work}

\input{methodology}

\section{Findings}

Across the five families, the consequences of using smart-home technologies varied with the relationship between a device, a child, and the family practices surrounding its use. A speaker could help a child request music and also become a source of frustration. A smartwatch could create room for independent movement yet remain difficult to wear. Cameras could reassure caregivers while prompting resistance from children. We organize these findings into four themes. The first examines how smart-home technologies provided support through familiar and adjustable environments and adapted forms of interaction. The second identifies sensory, communicative, and state-related mismatches between general-purpose devices and autistic children. The third examines the family work required to make these devices usable and govern their place in everyday life. The fourth shows how remote visibility enabled limited separation while keeping safety, privacy, and control under negotiation.

\subsection{Support Is Not Inherent: How Adaptation Turns Smart-Home Technologies into Support}

Smart-home technologies provided support when their functions could be fitted to a child's existing experiences, preferences, and ways of acting. Families created this fit in two areas. They used familiar and adjustable sensory conditions to respond to emotional states, and they adapted prompts or interaction channels to help children participate in everyday activities.

\subsubsection{Creating Sensory and Emotional Support Through Familiarity and Adjustability}

Familiarity helped caregivers identify sensory inputs that an individual child associated with safety and comfort. When P1's child experienced sensory overload, the caregiver first distinguished between withdrawal, when she wanted to remain undisturbed, and emotional breakdown, when she needed active reassurance. During the latter, the family drew on sounds that already carried meaning for her:

\begin{quote}
We create an environment that she feels is familiar and safe. For example, we wake the speaker and play music that is familiar and feels safe to her, sometimes even an ordinary piece of music that she once found cheerful or that carries a special memory.
\end{quote}

The family's knowledge of the child gave the music its supportive role. They knew which sounds she recognized, which felt safe, and which were connected to positive memories. P5 similarly explained that ``once repetitive or irritable emotions arise, music from the speaker can quickly calm him down and reduce episodes of breakdown and crying.'' Variation across children was also evident in ordinary-family Weibo posts. Six posts, WB-02 through WB-07, described children becoming frightened or crying in response to robot vacuums. In WB-01, however, a parent used a robot vacuum to help the child sleep: ``The baby needs white noise, so whenever he falls asleep, I turn on the robot vacuum.'' These contrasting responses show why families' knowledge of their own children was central to selecting supportive sensory inputs.

The adjustability of smart-home devices was also important because the same child could need different environmental conditions across activities and emotional states. P3 illustrated how the family used adjustable lighting to respond to these changes:

\begin{quote}
Smart lights can be adjusted to better fit different situations in a child's daily life. We use bright light for studying, warmer light for resting to reduce stimulation, and a night-light setting when she is emotionally unsettled. In our home, we usually change the scenes manually through voice commands because automatic adjustment can easily make mistakes.
\end{quote}

The different scenes allowed one lighting system to follow changes in the child's activities and emotional condition. Manual control also enabled the caregiver to correct a setting when automatic adjustment did not match the child's needs. P1 described a broader form of adjustment involving lighting, sound, airflow, and temperature. Through these adjustable functions, caregivers could translate changes in the child's condition into corresponding changes in the home environment. Guided by close knowledge of the child, familiar sensory inputs and flexible environmental controls could support emotional calming and recovery.

\subsubsection{Supporting Everyday Participation Through Adapted Prompts and Interaction Channels}

Families also adapted how devices presented an activity. P1's child resisted the directional reminders built into a smart toothbrush, although the family valued its prompting and timing functions. A technically skilled family member retained the brushing instructions and changed their form:

\begin{quote}
She does not like someone reminding her ``up, down, left, right,'' and becomes a little resistant. Someone in our family works in this area, so we changed the immediate prompts into a children's song. The song reminds her to brush upward and downward, and we try to loop it for the time she needs to brush, in order to reduce the difficulty and discomfort of brushing.
\end{quote}

The song carried the same information about movement and duration in a form selected for this child. The family located the source of resistance in the delivery of the instructions and redesigned that part of the activity. The evidence therefore supports a task-specific adaptation intended to make brushing more manageable.

Alternative interaction channels supported participation in a similar way. As P5 explained, ``Sometimes his aunt or grandparents cannot understand which song he wants, so he can speak to the speaker himself.'' The caregiver also observed greater willingness to initiate simple voice commands. When recognized, these brief requests gave the child a bounded route through which to act on a preference. P4's child used a smartwatch to call the caregiver, convey his immediate state, and add several friends. A Xiaohongshu post outside an explicit autism context, XHS-G01, described a child discussing loneliness with a familiar voice assistant after withholding those feelings from family members. The post illustrates how stable responses may open an additional channel for expression. Across these accounts, adapted prompts and interaction channels helped children enter specific activities and communicate particular intentions.

\subsection{When General-Purpose Technologies Misfit the Child: Sensory, Communicative, and State-Related Barriers}

General-purpose smart-home technologies are designed around assumptions about how users perceive prompts, produce commands, and understand feedback. The devices encountered in these families often expected sustained tolerance of wearable or auditory outputs, speech delivered in a recognizable form and sequence, and an ability to infer operating states from limited cues. Three types of mismatch became visible when these expectations met children's sensory experiences, expressions, and interpretations.

\subsubsection{Misfits in Sensory Intensity and Response Timing}

Children's smartwatches illustrate how a device's mode of access can limit an otherwise useful function. Location and communication depend on the watch remaining with the child, usually through continued wear. Four caregivers, P1 through P4, described difficulties with this condition. Some children found objects on the wrist or neck uncomfortable, while others resisted the change from their established routines. Calls and messages could also produce vibrations or prompts that became difficult to tolerate. P4 described an occasion when several calls made the watch vibrate repeatedly:

\begin{quote}
The watch kept vibrating, probably because I called several times. He did not know how to answer it, so he kept hitting it and eventually took it off and threw it onto the sofa. I do not call repeatedly now because I am afraid it will annoy him.
\end{quote}

Removing the watch ended the repeated vibration and also interrupted the location and communication connection. P3's child generally left the watch on a table because wearing objects on the wrist or neck was uncomfortable. P1 anticipated that her child might remove and lose a watch during sensory overload and considered attaching a button-style locator to clothing. Across these families, sensory compatibility directly shaped whether the device's safety functions remained available during everyday use.

Auditory outputs created similar barriers, although the evidence was less extensive. P3 described the speaker's replies as ``very long and slow'' and explained that the child ``basically cannot listen to the end,'' preventing a complete exchange. In the ASD-context Xiaohongshu post XHS-A01, the sound following an incorrect response intensified repeated experiences of failure with an educational robot. Turning off the sound later did not immediately restore use, and the family suspended the device until the child showed interest again. The ordinary-family post XHS-G02 also described a child being startled by unexpected voice announcements after home devices restarted. These cases show how intensity, duration, timing, and repetition can influence whether a general-purpose device remains tolerable and usable.

\subsubsection{Misfits Between Children's Expressions and Machine Interaction Rules}

Voice assistants organized interaction through rules governing recognizable speech, turn-taking, and error repair. Users had to speak clearly enough for the system, time a request around the wake response, and reformulate an utterance after a recognition error. P2, P3, and P5 all described occasions when a speaker misunderstood or failed to recognize a child's request, produced an irrelevant answer or the wrong song, and contributed to rising frustration or emotional escalation. The evidence was strongest around recognition and repair, where a child could express a clear intention and still receive an unrelated result.

P5's child sometimes varied in pronunciation and speaking rate when asking for music. The assistant then played the wrong content, and repeated requests did not reliably correct the error:

\begin{quote}
Several times he repeatedly called out the song he wanted, but it kept playing the wrong content. He would throw small toys, cover his ears, and lose his temper.
\end{quote}

The interaction continued to produce sound while withholding the expected song. Communication failure consequently developed into sensory and emotional distress. P2's child attempted to correct the speaker after it gave an irrelevant answer or selected the wrong song, yet the device misunderstood the correction and repeated the error. P3 likewise described poor speech recognition, inaccurate feedback, and occasions when the speaker became stuck or failed to respond. The child sometimes called the caregiver to report that the speaker could not understand or play the requested music. In both accounts, the intended request was present, while repeated recognition failures prolonged the exchange and made communication more difficult.

Other accounts extended this mismatch to the rules that precede recognition. The Weibo post WB-08 described a child who knew which song he wanted but had difficulty following the required sequence of saying the wake phrase, waiting for the assistant to respond, and then making the request. In the general-context Xiaohongshu post XHS-G03, interruption, partial recognition, and lengthy audio responses made interaction similarly difficult to sustain. P4 expected that a voice assistant would have difficulty recognizing the child's unclear speech and dialect, and the family therefore did not adopt one. Across these examples, restrictive language requirements, frequent recognition errors, and limited support for recovering from errors turned simple requests into difficult exchanges, sometimes ending in emotional distress or non-use.

\subsubsection{Misfits Between Children's Understanding and Device States}

Smart-home devices also relied on users to infer their internal states and the consequences of an action. Some devices conveyed this information through anthropomorphic language. P1 described how a robot vacuum announced that its tank needed attention:

\begin{quote}
Our robot vacuum used to say, ``Master, master, I have no water.'' The child could not immediately understand what it meant. She thought the robot needed a drink, so she kindly filled a cup with water and poured it onto the robot, causing a short circuit. This may be a low-probability event, but it can happen.
\end{quote}

The announcement invited a literal and social interpretation. The device presented a mechanical condition as the voice of a thirsty agent, while leaving the water tank and appropriate maintenance action unspecified. This ambiguity produced a concrete consequence when the child's caring response caused the short circuit. The case shows how humanlike wording can make a device engaging while obscuring the operational information required for safe action.

Other state cues created briefer versions of the same difficulty. On P1's induction cooker, an illuminated indicator did not clearly distinguish whether the appliance merely had power or had begun heating. As P1 explained, ``She may only see that a light is on, but she does not know which light it is.'' The family's face-recognition lock could also reopen when the child glanced back after closing the door, allowing her to walk away without noticing the state change. Across these cases, ambiguous indicators and silent automatic transitions made device behavior difficult to anticipate. A maintenance prompt could be acted on literally, a light did not clearly distinguish power from heating, and a lock could reverse a completed action without drawing attention to the change. These mismatches contributed to equipment damage and an unsecured door.

\subsection{Support Is Accomplished Through Family Work: Making Technology Usable and Governing Its Boundaries}

The mismatches described above frequently led to family intervention. Caregivers translated between children and machines, altered interfaces, configured household systems, and maintained access to services. They also managed the time, content, and intensity of device use. This work enabled many of the supportive outcomes described in the first theme and exposed the knowledge, time, and resources required to sustain them.

\subsubsection{Translating, Configuring, Modifying, and Maintaining General-Purpose Technologies}

Translation was one of the clearest forms of family work. When P2's child became frustrated after a speaker misunderstood him, the caregiver first helped him calm down, inferred his intended question, and considered how the assistant had interpreted it. She then reformulated the request:

\begin{quote}
I translate the child's question into something that will not make Xiaodu misunderstand it and that it can answer accurately. I know what the child wants to ask and what Xiaodu may have misunderstood, so I help translate between them.
\end{quote}

This translation required the caregiver to understand two communicative systems at once. She interpreted the child's intention and anticipated the wording the machine would accept while also managing the child's emotional response to failure.

Modification addressed mismatches that could not be resolved during a single exchange. P1's family rewrote the toothbrush prompts and was exploring a smart mirror that could convert weather information into clothing suggestions. P1 connected this work to a gap in available products: ``Products designed specifically for children are basically absent from the market, so sometimes we make things ourselves.'' Relevant technical expertise within the family made these adaptations possible. Families without comparable expertise may have few practical means to turn general-purpose smart-home products into devices that address their own autistic child's specific needs.

Configuration and maintenance required continuing work and financial investment around the devices. P5 retained physical buttons and adult control over more complex appliances, while P4's uncertainty about configuring a speaker contributed to the decision not to introduce one. In XHS-A01, continued use of an educational robot required parental accompaniment, changes to course content, contact with platform staff, and possible subscription payments after the free period. Another ASD-context Xiaohongshu parent, XHS-A02, returned a nearly 6,000-yuan robot after its screen use, repetition, and need for parental accompaniment outweighed its perceived value. The usability of these systems therefore extended beyond the purchased hardware to include family knowledge, time, money, and access to external support.

\subsubsection{Governing Time, Content, and Intensity at the Boundaries of Family Life}

Families also intervened when device use began to affect sleep, attention, or shared household space. P2's child raised the speaker to maximum volume, became increasingly excited through interaction, and wanted to continue watching stimulating short videos. The family responded by adjusting several boundaries:

\begin{quote}
He likes to turn the volume all the way up. We ask him to turn it down, and we have set a usage schedule so that he cannot use the speaker after eight in the evening. Interacting with it makes him excited, and he needs time to settle down before bed.
\end{quote}

The schedule addressed the timing and intensity of an interaction that the child found engaging. P2 also restricted phone use, removed access to Bilibili, and selected alternative content. P3 encountered a related household problem when the child's repeated songs and videos supported the consolidation of knowledge while disturbing other family members. These cases expanded the assessment of a device beyond the individual user to include its effects on family routines.

Other families governed access through dedicated controls, disconnection, or timers. P1 created a spare phone containing only smart-home controls, separating environmental functions from other applications. The restricted phone also became less attractive to the child, which reduced willingness to use it. Ordinary-family Weibo posts described unplugging a speaker after a child repeatedly controlled whole-home lighting in WB-11 and using assistant timers to end television time in WB-12 and WB-13. These interventions had varied effects. A timer could provide an acceptable stopping point, while stricter controls could weaken participation or end use entirely. Governing smart-home technologies involved repeated decisions about how a child's use could coexist with sleep, shared space, and continued human interaction.

\subsection{Monitored Independence: How Remote Visibility Reconfigures Safety, Letting Go, and Privacy}

Cameras, location watches, speakers, and connected alerts allowed caregivers and children to spend more time physically apart while retaining information and contact. This arrangement created a form of monitored independence. Children gained greater distance or time alone, and caregivers continued to observe, interpret, and respond. The same monitoring practices also raised questions about children's objections, parental responsibility, and the collection of household data.

\subsubsection{Supporting Limited Separation Through Visibility and Contact}

Location and communication technologies created room for separation when children wanted greater independence. P4's child strongly insisted on going out alone and, when he noticed his mother following, would shout, ``Mom, don't follow me!'' His limited traffic awareness and tendency to detour towards car parks continued to concern her. The location watch allowed the caregiver to remain farther away while retaining an approximate location and a way to make contact. The ASD-context Xiaohongshu post XHS-A03 described a similar arrangement. A five-year-old who wanted to go out alone could use a watch to report buying ice cream or riding a bus, while the parent followed his location at a distance. In both accounts, remote information allowed children to move farther from visible parental presence while caregivers remained connected.

Monitoring also supported brief periods in which caregivers left the home. P1 described going to a film after the child had fallen asleep and receiving an alert when the child left the bedroom:

\begin{quote}
The device notified us, ``The child has come out of the bedroom; the child is awake.'' I contacted her immediately, first letting her hear my voice and then see my image, and then I drove home.
\end{quote}

The alert shortened the time before the caregiver became aware of the change. Audio and video restored contact, and the caregiver then judged that returning home was necessary. Remote visibility worked through this sequence of detection, interpretation, communication, and intervention. P5 similarly used camera images and speaker audio to assess the child's condition, while P3 described a watch as a channel through which the child could seek comfort during distress. Ordinary-family Weibo posts provided contextual examples: WB-14 described improvised contact through two speakers, WB-15 described a caregiver returning after seeing a child cry on camera, and WB-16 described parents following a child's coordinates until the child entered the home.

However, this separation remained limited. The alert and remote contact helped P1 respond before returning, while the child's need for care continued beyond what audio and video could provide. Remote visibility supported a limited period of separation because the caregiver remained attentive and physically available to intervene.

\subsubsection{Negotiating Children's Objections and Privacy Under a Safety-First Logic}

Although remote visibility helped caregivers manage safety during separation, children sometimes resisted being followed, watched, or recorded. Their objections often appeared through behavior and use of space. P4's child protested when his mother followed him, disliked an adult remaining near the bathroom, and sought solitude when upset. P5 described a direct response to a domestic camera:

\begin{quote}
When he notices that the camera is pointing at him, he deliberately turns his back, covers it, or demands that I turn it off. He cannot fully explain his objection, but his behavior clearly rejects the camera facing him.
\end{quote}

By turning away, covering the lens, and asking for the camera to be switched off, the child expressed a boundary against monitoring. The ordinary-family Weibo post WB-17 described a child similarly blocking a camera.

Caregivers held different understandings of privacy and data collection. P5 worried that platforms might retain or leak location routes, household conversations, and recordings of emotional distress: ``I do worry about data storage and use, and I am afraid that platforms may retain or leak home audio and video.'' P1 regarded data authorization as an unavoidable condition attached to device use. P2 had never considered that a speaker might record domestic audio. These accounts ranged from explicit concern about exposure to limited prior awareness or resigned acceptance of platform practices.

Safety responsibilities made monitoring more difficult to relinquish. P4 explained, ``I know that when the child wears the watch, he is being located all the time, but without it I do not feel at ease. He has little awareness of danger when crossing roads, and sometimes walks straight towards something he likes.'' Her account shows why continuous location remained difficult to give up even when its intrusiveness was recognized. Families therefore adopted practical compromises. P4 stated, ``My bottom line is that he wears the watch only when going out or attending school. At home, if he does not want to wear it, I generally do not force him. We do not place cameras in the bedroom or bathroom.'' P5 treated cameras as part of a safety arrangement and temporarily turned one off when the child clearly objected. P3 kept cameras covered or disabled and activated them when monitoring became necessary. Across the five households, monitoring boundaries remained adjustable and unsettled. Children's objections influenced everyday practices, while caregivers' safety responsibilities and limited control over platform data continued to shape how far monitoring could be reduced.

\section{Discussion}

This study examined how families of autistic children encounter, adapt, and govern general-purpose smart-home technologies in everyday life. The findings show that the supportive value of general-purpose smart-home technologies varied across children, situations, and family arrangements. It emerged from relations among device behavior, children's sensory and communicative experiences and changing needs, and the family practices that adapted and governed use.
 This account connects four observations: families adapt ordinary devices to create sensory, emotional, and practical support; the same devices become barriers when their assumptions about bodies, speech, timing, or system states do not fit the child; supportive use depends on continuing family work; and remote monitoring produces a conditional form of independence shaped by safety and privacy negotiations. Below, we discuss how these findings extend prior work on autism and assistive technology, domestic technology and care, and children's agency and privacy.

\begin{figure*}[t]
    \centering
    \includegraphics[width=\textwidth]{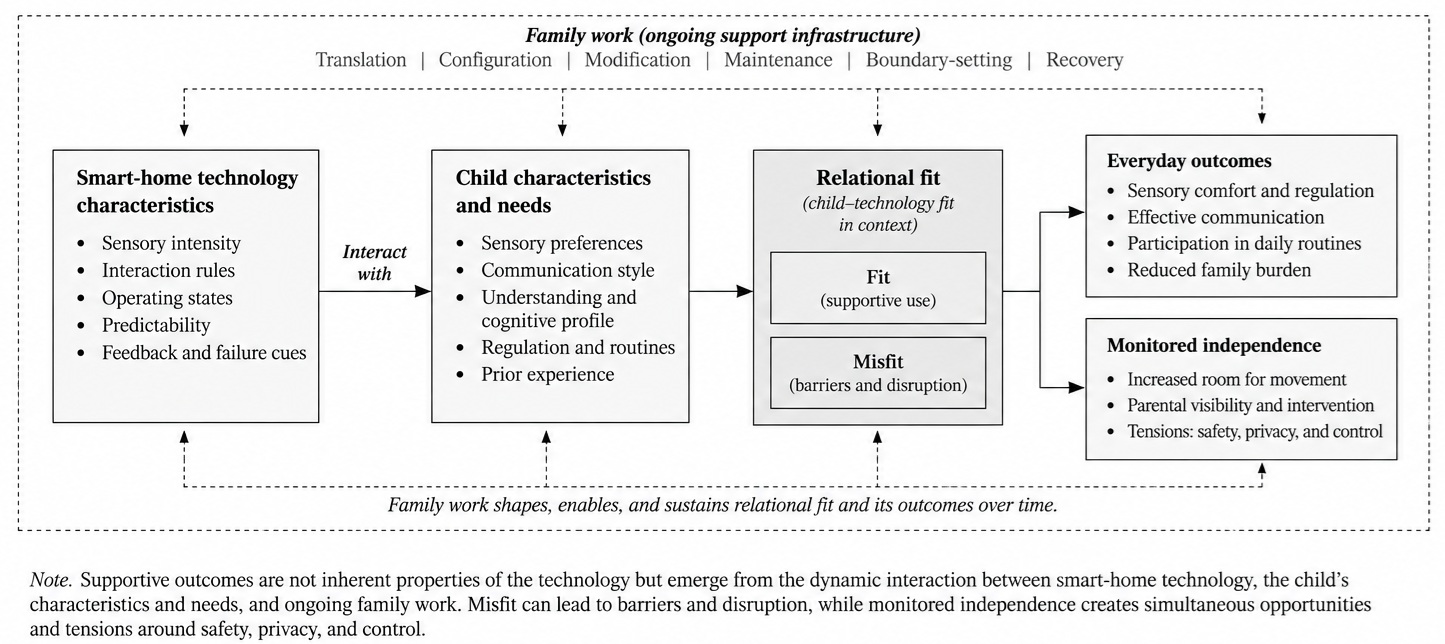}
    \caption{A conceptual model of relational fit, family work, and monitored 
    independence in general-purpose smart homes for autistic children.}
    \Description{A conceptual model showing how smart-home technology 
    characteristics interact with a child's characteristics and needs to 
    produce relational fit or misfit. Ongoing family work shapes and sustains 
    this relationship, which influences everyday outcomes and monitored 
    independence.}
    \label{fig:crux}
\end{figure*}

As seen in Figure~\ref{fig:crux}, we conceptualize support not as an inherent property of smart-home technology, but as an outcome of relational fit among the characteristics of the technology, the child's characteristics and needs, and the family work surrounding its use. When these elements align, technology can support sensory regulation, communication, and participation in everyday routines. When they do not, the same technologies can produce sensory, communicative, and state-related barriers. Family work is central to this relationship because caregivers continually translate, configure, modify, maintain, and govern technologies to sustain their fit with the child. Monitored independence further illustrates the relational nature of support: greater room for movement and action can coexist with parental visibility and intervention, creating continuing tensions among independence, safety, privacy, and control.

\subsection{Relational Fit in General-Purpose Smart Homes}

Prior autism technology research has questioned whose goals define technological support and has called for closer attention to autistic children's interests and agency~\cite{spiel2019agency}. Studies of technology use at home have likewise shown that families accommodate children's individual needs and manage perceived benefits and risks~\cite{laurie2019parental}, while deployments of purpose-built robots and applications emphasize personalization and sustained engagement~\cite{clabaugh2019robot,rechowicz2021aspect,kim2023routineaid}. Research on supportive environments further identifies light, sound, temperature, movement, and environmental structure as consequential design dimensions~\cite{wohofsky2023senshome,resch2026smarter,park2024masksound}. AACessTalk shows how parent-facing guidance and child-facing supports can restructure a specific communicative activity~\cite{choi2025aaccesstalk}, while caregiver responses to an AI home-hub concept identify anticipated forms of household support~\cite{perry2026assistive}. Our findings complement this work by tracing observed uses of existing consumer devices across multiple routines and by distinguishing task-specific support from broader claims about independence.

We describe this process as relational fit. A device became supportive when its outputs could be aligned with a child's existing preferences, tolerances, forms of expression, and ways of understanding action. Familiar music and adjustable lighting could help families respond to distress; an adapted brushing song recast a routine prompt in a child-specific form; and a watch or speaker could provide a usable channel for a specific request or contact. The nominal capabilities of playing music, controlling lights, issuing reminders, or making calls did not determine these outcomes on their own. Their effects depended on whether those capabilities were delivered in a form that the child could tolerate, interpret, and use in a particular situation.

Relational fit also explains why the same technology could alternate between support and disruption. Voice output could calm a child when it delivered familiar music and intensify distress when it repeatedly returned the wrong content. A wearable could support contact and location awareness while its physical feel or vibration made continued wear difficult. A lock intended to improve security could introduce risk when its recognition behavior or automatic state changes were not apparent. The value of a device was therefore dynamic: it changed with the child's state, the interaction history, the household setting, and the family's capacity to intervene. Personalization therefore involved an ongoing process of noticing responses, modifying arrangements, and deciding when a feature should be used, restricted, or abandoned.

The three forms of misfit identified in our findings make visible assumptions embedded in general-purpose products. Sensory and temporal misfit appears when a device assumes that vibration, loudness, delay, repetition, or abrupt output will be broadly tolerable. Communicative misfit appears when voice systems assume clear pronunciation, standardized language, orderly turn-taking, and the ability to repair errors according to machine-defined rules. State-related misfit appears when products assume that users can infer maintenance needs, operating modes, automation logic, and physical consequences from sparse or anthropomorphic feedback. Locating breakdown in these relations avoids treating the child as the sole site of difficulty. The problem can lie in the interaction between an individual child and a product built around a narrow image of the competent user.

Research on smart speakers and family life has already shown that connected devices participate in parent-child collaboration, conflict, discipline, and routine~\cite{beneteau2020alexa,garg2020smartspeakers}. Our cross-source comparison adds an important qualification. Sudden sounds, household disruption, and disagreements over control also appeared in general-family social-media posts. These problems also affected families outside autism contexts. The caregiver interviews show, however, that recognition failures and repeated outputs could develop into sensory and emotional distress in particular interactions. This distinction matters because it neither exceptionalizes autistic families nor assumes that a generic family design is sufficient. It directs attention toward flexible systems whose behavior can be made intelligible and tolerable across diverse users and changing situations.

\subsection{Family Work as the Infrastructure of Support}

The findings also extend scholarship on the labor that sustains domestic technologies. Prior work describes the configuration, troubleshooting, and coordination of networked devices as digital housekeeping~\cite{tolmie2007digitalhousekeeping}. It shows how technical authority may concentrate in household ``pilot'' users~\cite{koshy2021passenger}, how incomplete knowledge of platforms and data flows complicates configuration~\cite{abdi2019assistants}, and how protection, responsibility, affection, and anticipation make technical maintenance a form of domestic care~\cite{strengers2019protection,gruhlich2025digitalcare,frampton2026domestic}. Our study demonstrates how this work changes when general-purpose technologies are used around an autistic child.

Families performed work at several interconnected levels. They interpreted children's behavior and sensory states, translated between children's expressions and machine-recognizable language, configured scenes and permissions, altered prompts, selected alternative device forms, maintained connections, and decided who could operate which functions. They also governed volume, repetition, content, timing, and access after a device became attractive or useful. These activities were not merely technical tasks surrounding an otherwise autonomous product. They formed part of the mechanism through which the product acquired its supportive capacity. A voice assistant appeared responsive after a caregiver reformulated the child's request; a smart toothbrush became usable after its prompt was redesigned; and remote sensing became actionable only when a caregiver interpreted an alert and decided how to respond.

This finding complicates simple claims that automation reduces care work. Smart-home technologies sometimes saved time, extended attention, or enabled temporary separation, but they could also relocate work. Directly performing an activity might be replaced by anticipating system behavior, configuring acceptable outputs, monitoring use, repairing failures, or regulating intensity. The relevant question is therefore how care is reorganized and which forms of work become necessary, visible, or concentrated in particular family members. Families did not invariably describe this work as burdensome. Adaptation could be an expression of expertise and care, and some arrangements produced clear practical value. Nevertheless, benefits depended unevenly on access to time, technical knowledge, compatible devices, and responsive platform support. A household able to modify prompts or configure child-specific scenes had options unavailable to a caregiver who found configuration inaccessible or feared unpredictable output.

This dependence creates a risk that the apparent intelligence of a system is sustained by invisible human compensation. When a caregiver repeatedly translates speech, manually corrects automation, or curates every acceptable interaction, the family becomes a compatibility layer between the child and the product. Treating this labor as part of system performance changes how smart-home technologies should be evaluated. Usability assessments should account for setup, interpretation, repair, supervision, and boundary-setting over time alongside whether a child can complete an isolated interaction. They should also ask whose labor makes successful use possible and whether the arrangement broadens or narrows the child's own participation.

\subsection{Monitored Independence and Conditional Privacy}

Our findings further develop existing accounts of family surveillance. Prior studies show that parents weigh reassurance and protection against trust and children's self-direction when considering location tracking~\cite{vasalou2012nonadoption}, and that adults and young people may interpret records from locks, cameras, and other connected systems differently~\cite{ur2014intruders}. Work on multi-user smart homes similarly shows that installation and administrative control shape household participation~\cite{geeng2019control,zeng2019multiuser}. In our data, monitoring reorganized the relationship between distance and care.

Our findings show how monitored independence operated in everyday family life: children gained physical space while caregivers retained informational proximity and a capacity to intervene. Location watches, cameras, alerts, calls, and audio channels allowed some parents to follow from farther away, leave a child alone for a limited period, or respond after noticing distress. Independence in these cases was graded rather than absolute. The child could move beyond immediate adult presence, while the caregiver remained connected through data and prepared to act. This extends discussions of tracking beyond an opposition between surveillance and freedom: monitoring could enable a form of separation that a family might otherwise consider unsafe, while also preserving adult authority over where, when, and how that separation occurred.

Privacy remained central to this arrangement. Research has shown that autistic young people can articulate privacy concerns and use protective strategies~\cite{rocheleau2022privacy,yuan2023redcapes}, while work with Chinese parents of autistic children highlights contextual judgments about disclosure, protection, stigma, and access to support~\cite{wang2025chineseprivacy}. Our caregiver accounts add that privacy claims may also be interpreted through embodied and spatial behavior. Turning away from a camera, covering its lens, resisting observation near a bathroom, or seeking solitude can communicate a boundary against monitoring. Because our evidence comes primarily through caregivers, we cannot equate these interpretations with children's direct accounts. They nevertheless show that whether a privacy claim becomes legible within the household depends partly on adults recognizing behavior as communicative.

The force of these claims was conditional under a safety-first logic. Caregivers often narrowed monitoring by time, place, or purpose, yet retained final authority when they judged the risk of wandering, traffic, distress, or inability to seek help to be high. Privacy was therefore not settled by a single consent decision at installation. It was negotiated repeatedly as circumstances and perceived risks changed. Differences in caregivers' awareness and concerns further complicated this negotiation. The longer-term storage of domestic data and institutional access to it remained unsettled within these households. Monitored independence thus names both an enabling arrangement and an unresolved power relation: it expands movement through continued visibility, while leaving decisions about observation and intervention unequally distributed.

\subsection{Design Implications}

These findings suggest that smart-home design for neurodiverse families should prioritize adjustable fit over a fixed package of autism-specific functions. Support should be configurable across sensory intensity, response length, timing, modality, repetition, and physical form. Families should be able to create child-specific scenes, preserve familiar content, set narrow ranges for volume or brightness, and choose whether an action is automatic, suggested, or manually triggered. Such flexibility should remain understandable without specialist technical expertise. Because children's preferences and tolerances can change across activities and states, profiles should be easy to revise rather than treated as permanent diagnostic settings.

Systems should also fail in ways that are legible, low in stimulation, and recoverable. The findings do not establish recoverability as an observed household outcome; instead, repeated misrecognition, opaque device states, and accumulated frustration make it a design requirement. Voice assistants could confirm an interpreted request before acting, offer brief responses, display or speak a small number of clear alternatives, and allow users to repeat, point, tap, or select rather than requiring increasingly precise reformulation. Device messages should clearly name the affected component, explain the required action, and indicate the device's current operating state. Systems should also make stopping or undoing an automation immediate. Error feedback should avoid abrupt sounds, escalating repetition, or lengthy dialogue. After a breakdown, the system should provide a predictable route back to a known state instead of requiring the child or caregiver to reconstruct what happened.

Design should reduce family translation and maintenance work without eliminating family participation. Caregivers need accessible tools for editing prompts, assigning functions, reviewing recent actions, setting schedules, and choosing alternative input channels. Reusable configurations, clear permission models, local explanations of automation, and dependable technical support could reduce repeated setup and troubleshooting. At the same time, systems should not assume that the caregiver will permanently mediate every interaction. Following calls for meaningful participation with autistic children~\cite{frauenberger2019outsidethebox,maun2024methods}, customization should include ways for children to express preferences through accessible choices, demonstrations, or behavioral responses. Customization should give children meaningful options while allowing caregivers to set context-dependent boundaries.

Monitoring features should support graduated, negotiable independence. Families need controls scoped by location, time, data type, and purpose: for example, location sharing during travel without continuous tracking at home, event-based alerts instead of persistent streaming, and explicit camera-free spaces. Child-facing indicators should make monitoring visible, and systems should provide accessible ways to request a pause, turn away a sensor, or signal discomfort. When caregivers override such preferences for immediate safety, the override should be limited, visible, and reviewable. Multi-user controls and context-sensitive privacy mechanisms proposed in earlier work~\cite{zeng2019multiuser,liu2025privacycontrols} are especially relevant here, but they should be designed around changing care relationships rather than static owner and guest roles.

Finally, evaluation should move beyond short demonstrations of feature accuracy. Longitudinal assessment should examine whether a device remains tolerable, how errors accumulate, how much work is required to sustain it, how it affects other household members, and whether children gain meaningful options for communication, participation, or movement. This household-level perspective also guards against interpreting caregiver convenience as equivalent to child benefit, or increased device use as equivalent to greater independence. A successful system is one whose benefits remain usable without requiring continuous hidden compensation and whose arrangements can be renegotiated as the child, family, and technology change.

Taken together, our findings position smart-home support as a contingent and negotiated household accomplishment. General-purpose technologies became valuable resources through relations of fit, family labor, and continuing governance. Designing for these relations shifts attention away from isolated device capabilities and toward the conditions under which children and families can understand, adjust, refuse, repair, and reshape domestic technology.

\subsection{Limitations and Future Work}

Our findings should be interpreted in light of several limitations. First, the interview data came from five caregivers and therefore capture caregiver accounts of children's experiences rather than children's direct perspectives. The sample cannot represent the heterogeneity of autistic children's sensory preferences, communication styles, support needs, family structures, and access to technical resources. Caregivers' interpretations were valuable for understanding household decisions and care practices, but may not fully reflect how children understood or evaluated the same technologies. Future work should engage autistic children directly through accessible and flexible participatory methods and include other household members whose responsibilities and authority may differ.

Second, the social-media datasets were shaped by platform affordances, keyword and hashtag practices, public self-presentation, and our screening decisions. Autism-context Xiaohongshu posts provided naturalistic accounts, but users' diagnostic descriptions could not be independently verified. Weibo posts were collected using child- and smart-home-related keywords and were used only as general-family context, rather than as evidence about autistic children. Consequently, the number of posts associated with a theme should not be interpreted as indicating its prevalence. Future studies could compare additional platforms and combine public posts with participant diaries or other recruited data to examine experiences that families may not share publicly.

Third, our data were primarily retrospective and cross-sectional. They cannot establish how support, breakdown, family work, or privacy negotiations develop as children's needs, household routines, and technologies change. Longitudinal in-home research could follow the adoption, abandonment, and reconfiguration of devices, particularly after repeated failures or changes in children's circumstances. The analysis also involved translating Chinese materials into English. Although interpretations and quotations were repeatedly checked against the source materials, some linguistic nuance or analytic emphasis may remain. Future research could incorporate additional bilingual review and researcher perspectives while documenting how AI assistance shapes qualitative interpretation.

\section{Conclusion}

This study examined how families of autistic children make general-purpose smart-home technologies usable in everyday life. Across caregiver interviews and social media data, we found that these technologies do not provide support through functionality alone. Their value depends on relational fit, i.e., whether sensory outputs, interaction rules, and operating states align with a child’s needs and family routines. Achieving this fit requires continuing family work, including translation, configuration, modification, maintenance, and boundary-setting. Remote monitoring can also enable monitored independence, allowing children greater freedom while caregivers retain oversight, but it introduces persistent tensions among safety, privacy, agency, and control. These findings shift accessible smart-home design from adding isolated features toward supporting adjustable interaction, legible and recoverable failures, lower-burden configuration, and negotiable monitoring arrangements. Ultimately, smart homes become supportive not when they maximize automation, but when children and families can understand, adapt, refuse, repair, and renegotiate how these technologies participate in domestic life.

\bibliographystyle{ACM-Reference-Format}
\bibliography{references}

\appendix

\end{document}

%% file: introduction.tex
\section{Introduction}
\label{sec:introduction}

In many households, smart-home technologies are introduced as tools for convenience, automation, and safety. In households with autistic children, however, these technologies can intersect more directly with sensory regulation, communication, daily routines, and caregiver oversight. Autistic children may differ in their responses to sound, light, verbal prompts, unexpected changes, and other environmental stimuli, while some families rely on carefully structured routines and individualized strategies to support participation in everyday activities~\cite{sun2024unfulfilled,wohofsky2023senshome}. As a result, a feature that functions as a minor convenience in one household may become an important support, or a source of disruption, when used by an autistic child. This makes the home an important setting for examining how general-purpose smart technologies fit the needs of autistic children and the caregiving practices that surround their use.

Smart-home products are increasingly becoming a source of support in family daily life. Smart speakers/displays with Voice assistants can play users' preferred music or set routines, smart cameras and location devices promise safety and reassure when a caregiver cannot remain in the same room, and connected locks, lights, and appliances can be adjusted remotely. For families of autistic children, these promises are especially salient and consequential because household environments may need to accommodate sensory sensitivities, changing states of regulation, and the practical work of daily routines, safety, and care~\cite{sun2024unfulfilled,wohofsky2023senshome}. What families describe, however, is a conditional relationship. The same device can be reassuring in one family and frightening in another, useful for one task and burdensome for another, and supportive only when its output, interaction rules, and operating state fit a child's needs and the family's arrangements. Instead of being just an edge case, these differences shape whether consumer smart-home technologies become workable parts of family life or introduce additional uncertainty and caregiving work.

Prior HCI (Human-Computer Interaction) research has approached these questions through two largely separate lines of work. One examines assistive and participatory technologies for autistic people, including socially assistive robots, routine support, and methods for engaging autistic children in design~\cite{clabaugh2019robot,kim2023routineaid,frauenberger2019outsidethebox,maun2024methods}. This work has drawn attention to children's agency, contextual fit, and the limitations of treating technology as a substitute for human relationships~\cite{spiel2019agency}. Recent ethnographic work has also examined how young autistic children use digital technologies independently and with family members at home~\cite{silva2026exploring}, but connected smart-home ecosystems remain a distinct and less examined part of domestic life. Another line of research studies smart homes as sites of family dynamics, digital housekeeping, child safety, and privacy negotiation~\cite{beneteau2020alexa,garg2020smartspeakers,tolmie2007digitalhousekeeping,frampton2026domestic,sun2021childsafety,zeng2017privacy,wang2025chineseprivacy}. However, we still know comparatively little about how commercially available smart-home ecosystems are incorporated into households with autistic children.  Research on autism and technology often centers bespoke systems, clinical goals, or single devices, while research on smart homes usually reports general-parenting or multi-user contexts in which children's sensory, communicative, and agency-related needs are not the analytical focus.

This gap leaves two aspects of everyday smart-home use insufficiently understood.  First, support is usually analyzed at the level of an individual device or feature, even though families live with multiple devices as part of an assembled environment that they continually configure and adjust.  Second, valued functions such as independence, routine support, or safety are often separated from their cost and associated family work, including translation and configuration labor, maintenance, monitoring, and privacy negotiation. Product-centered evaluations can obscure this work, even when it strongly shapes whether a technology remains usable in everyday caregiving.  In this sense, we treat recoverability not as an already established household condition but as a design issue raised by mismatched outputs, opaque device states, and the repair work families perform. As such, in this study, we seek to answer the following three research questions:

\begin{itemize}
\item \textbf{RQ1.} Under what conditions do general-purpose smart-home technologies support or hinder autistic children's sensory regulation and everyday participation?
\item \textbf{RQ2.} What adaptation, translation, configuration, maintenance, and boundary-setting work do families undertake to make and keep these technologies usable in everyday family life?
\item \textbf{RQ3.} How does remote monitoring reconfigure children's independence, safety, privacy, and control within the household?
\end{itemize}

\begin{figure*}[t]
    \centering
    \includegraphics[width=0.8\textwidth]{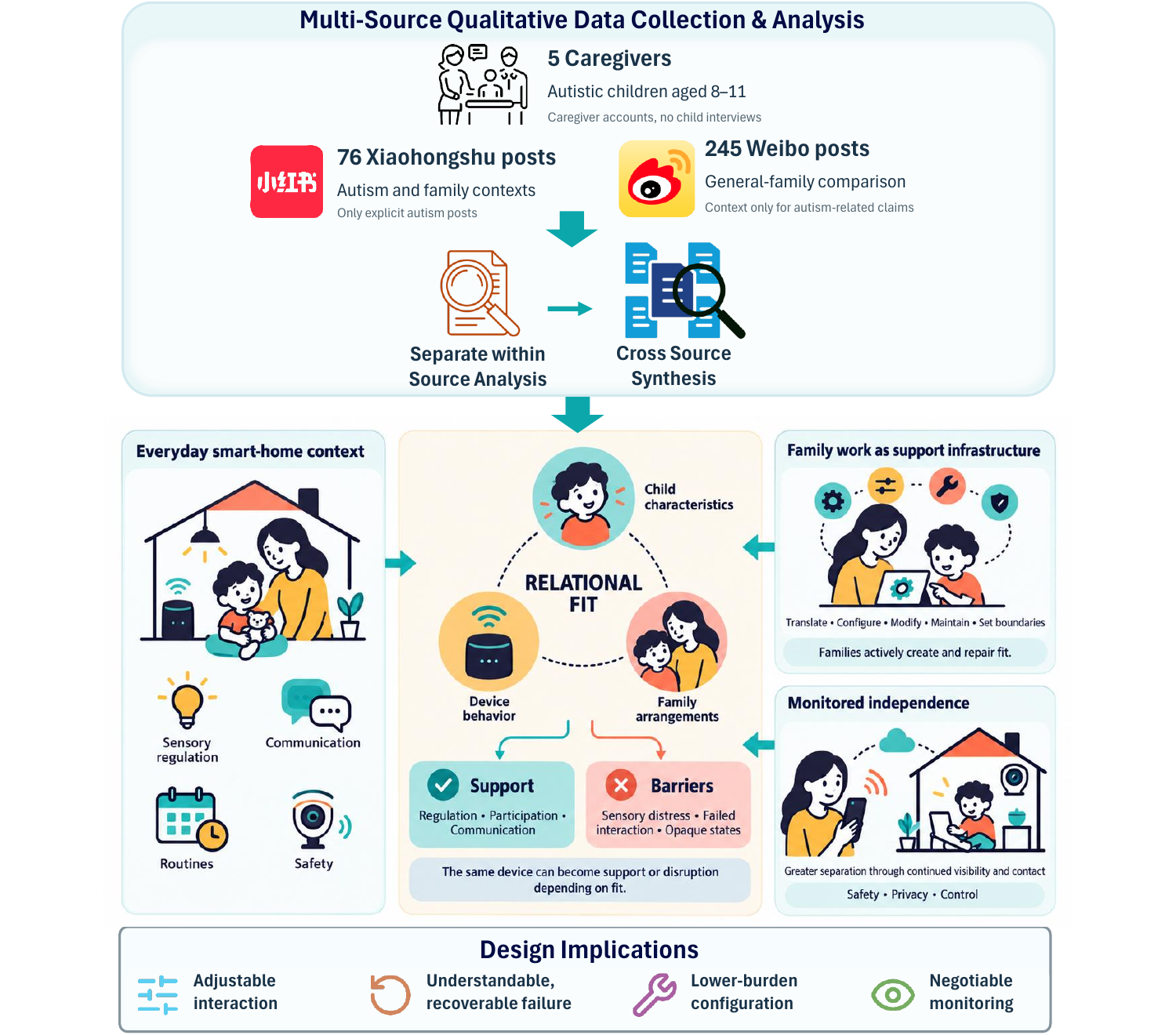}
    \caption{Graphical abstract of how general-purpose smart-home technologies become supportive for autistic children and their families.}
    \Description{Graphical abstract summarizing the study's conceptual findings. An everyday smart-home context containing an autistic child, caregiver, and connected devices leads to a central relational-fit framework linking child characteristics, device behavior, and family arrangements. When these elements align, smart-home technologies can support regulation, participation, and communication; when they do not align, they can contribute to sensory distress, failed interactions, and difficulties understanding device states. A family-work panel shows caregivers translating, configuring, modifying, maintaining, and setting boundaries around technologies in order to create or restore fit. A monitored-independence panel shows a caregiver remotely connected to a child at home, illustrating how smart-home monitoring can enable greater separation while maintaining visibility and contact, with ongoing tensions among safety, privacy, and control. A final design-implications panel highlights adjustable interaction, recoverable failure, lower-burden configuration, and negotiable monitoring.}
    \label{fig:graphical-abstract}
\end{figure*}

To answer these questions, we used a multi-source qualitative design as shown in Figure \ref{fig:graphical-abstract}. We conducted five interviews with caregivers of autistic children and analyzed them together with personal-experience posts from Xiaohongshu and Weibo. The three corpora were analyzed separately and then compared. Interviews served as the core explanatory evidence for understanding family practices, decisions, and experiences. Xiaohongshu included posts that explicitly situated experiences in autism-related contexts, which provided naturalistic autism-related evidence, as well as posts without an explicit autism-related context, which were used as adjacent or supplemental general-family material. Weibo was used solely as general-family context and comparison and was not treated as evidence about autistic children. The data roles were determined by how directly each source could address the research questions rather than by corpus size.

Our analysis produced four cross-source themes (Figure \ref{fig:graphical-abstract}). First, families turned ordinary devices into support by adapting environmental outputs, prompts, and interaction channels to the child. Second, general-purpose technologies became barriers when their sensory intensity, communicative rules, or operating states did not fit the child. Third, supportive use depended on family work, including translation, configuration, modification, maintenance, and boundary-setting. Fourth, remote monitoring enabled a conditional form of independence while preserving caregiver oversight and generating continuing negotiations over safety and privacy. Across these themes, the value of smart-home technology depended less on the amount of automation it provided than on how well families could adapt, understand, and manage it. This shifts attention toward whether a technology can be adjusted to fit the child, whether failures are legible and recoverable, how much family work the technology creates or redistributes, and whether monitoring arrangements can be renegotiated as children's needs and risks change.

This paper makes three contributions. Empirically, it provides a multi-source account of how families of autistic children appropriate general-purpose smart-home technologies and encounter breakdowns, while differentiating the evidentiary roles of interviews, Xiaohongshu posts, and Weibo posts. Conceptually, it develops relational fit to explain why the same device can shift between support and disruption, treats family work as part of the infrastructure through which technological support is produced, and theorizes monitored independence as a conditional arrangement among distance, safety, privacy, and control. For smart home design, it identifies requirements for adjustable fit, legible and low-stimulation failure, recoverable interaction, lower-burden family configuration and maintenance, and graduated, negotiable monitoring. These contributions are interpretive and bounded; we do not claim prevalence, universal effects, or clinical diagnoses.

%% file: related-work.tex

\section{Background and Related Work}

We draw on four literatures. Work on autism-related technologies and supportive environments establishes the importance of individual needs and environmental conditions. Studies of children's agency and privacy and of domestic care and maintenance explain shared use, household authority, and the work of sustaining technology. Together, these literatures suggest that support is not an inherent device property but a situated and ongoing accomplishment. We therefore distinguish deployed systems from design concepts and caregiver expectations, as well as direct child perspectives from caregiver- or observer-reported accounts. We use \emph{smart-home ecosystems} to mean home-centered arrangements of connected consumer devices. Fixed household systems are central; portable devices such as location wearables are included when families configure them as part of domestic care, safety, and control. This boundary follows the role a device plays in family routines while retaining the home as the analytic center.

\subsection{Autism, Assistive Technology, and Everyday Support at Home}
\label{sec:rw-asd-support}

Technology research involving autistic children raises a basic question: whose goals define support? A critical review by \citet{spiel2019agency} finds that many systems prioritize externally defined behavioral, educational, social, or therapeutic goals over children's own interests and agency. This distinction is especially important at home, where technologies become part of relationships, sensory environments, and household routines. Support should therefore be assessed not only by whether a task is completed, but also by whether the technology fits a child's sensory preferences, communication practices, interests, and existing ways of acting.

Evidence from in-home and everyday settings shows the importance of adaptation and personalization. A multinational parent survey documents how families accommodate autistic children's technology use within perceived benefits, risks, and household constraints~\cite{laurie2019parental}. A longitudinal in-home robot deployment likewise foregrounds personalization and sustained engagement~\cite{clabaugh2019robot}. \citet{rechowicz2021aspect} evaluate an Alexa-based system for echoic-skill practice with nine autistic children, whereas \citet{kim2023routineaid} study a mobile application that externalizes routine-support elements for autistic adults. Recent communication work further suggests that support may require redesigning interactional structures rather than expecting autistic children to adapt to existing interfaces: in a two-week home deployment with 11 minimally verbal autistic child--parent dyads, AACessTalk combined parent-facing conversational guidance with child-facing contextual vocabulary cards to structure reciprocal turn-taking~\cite{choi2025aaccesstalk}. Caregiver responses to an AI home-hub concept identify expected support for daily activities and caregivers' own needs, but do not establish outcomes from children's use~\cite{perry2026assistive}. Taken together, this work foregrounds fit, scaffolding, and sustained personalization, but its evidence is distributed across bespoke systems, single activities, adult participants, therapeutic or educational contexts, and caregiver expectations. It tells us less about how general-purpose consumer ecosystems fit the changing conditions of family life, or how a device may scaffold a particular act without producing broad gains in independence.

\subsection{Consumer Smart Homes as Supportive Domestic Environments}
\label{sec:rw-smarthome-support}

Research on supportive environments shifts attention from individual devices to domestic conditions. The SENSHOME project relates light, sound, temperature, air quality, and movement to requirements for safety, comfort, accessible interfaces, and daily structure for autistic people~\cite{wohofsky2023senshome}. \citet{resch2026smarter} translate this perspective into guidelines for commercially available devices, including voice-assistant schedules, alarms, presence sensors, and environmental controls. At a more focused level, \citet{park2024masksound} explore sound masking as a response to noise sensitivity among autistic people. This work establishes the relevance of adjustable sensory conditions. However, requirements, guidelines, and focused interventions provide less evidence about how families appropriate consumer products over time, or what allows a routine to remain predictable and recoverable when an automation, prompt, or sensory output is inappropriate.

Broader family research adds the social conditions of use. Smart-speaker studies describe parent--child collaboration, discipline, conflict, independent activity, and differences in family routines and perceptions~\cite{beneteau2020alexa,garg2020smartspeakers}. Parents' concerns span physical safety, content, digital security, and privacy as children become active users~\cite{sun2021childsafety}; smart-home users more generally interpret risk through household practices as well as technical features~\cite{zeng2017privacy}. Although these studies are not autism-specific, they show that value and acceptability emerge within shared routines and relationships. Neither research stream, however, centers the assumptions that general-purpose products make about sensory tolerance, speech, response timing, or users' understanding of system states. The resulting gap concerns interactional fit: whether families can anticipate what a system will do, interpret its state, interrupt unsuitable output, repair a failed exchange, and restore a routine after breakdown.

\subsection{Children's Agency, Privacy, Safety, and Multi-User Control}
\label{sec:rw-children-agency-privacy}

Shared use raises the question of who can shape a household's technological arrangements. Participatory work with autistic children demonstrates individualized ways of designing for positive and meaningful experiences, while methodological reviews emphasize accessible and flexible engagement~\cite{frauenberger2019outsidethebox,maun2024methods}. \citet{morris2025chromaconnect} similarly co-designed and evaluated a social-play prototype with a neurodiverse group of children, showing how technology can make divergent play modes and nonverbal expressions of social-play initiation visible. Privacy research likewise challenges assumptions that autistic young people cannot understand or express privacy preferences. Autistic teenagers describe protective strategies and concerns about online risks~\cite{rocheleau2022privacy}, and autistic and non-autistic children engage with contextual privacy scenarios through the RedCapes game~\cite{yuan2023redcapes}. These studies establish the importance of children's perspectives. They leave less understood how those perspectives are expressed or mediated when domestic sensing makes privacy a matter of rooms, bodies, cameras, location traces, and withdrawal from interaction---especially when a child does not express an objection in abstract privacy language.

In smart homes, meaningful participation also depends on how control is distributed. Device selectors and installers often retain administrative and repair authority, producing cooperation as well as tension~\cite{geeng2019control}; multi-user access, notification, and intelligibility mechanisms can broaden participation~\cite{zeng2019multiuser}. Children have proposed awareness cues, stronger authentication, and context-sensitive controls~\cite{liu2025privacycontrols}, yet product marketing often leaves child-safety features and data practices unclear~\cite{sun2024unfulfilled}. Speculative work with families further shows that smart-home datafication raises questions about who can see household data, who can act on it, and how family members imagine future control~\cite{wang2025famidata}. Studies of family surveillance expose a related tension: parents and teenagers differ over photographic and text records from connected locks and cameras~\cite{ur2014intruders}, while parents considering location tracking weigh safety and reassurance against trust and children's self-direction~\cite{vasalou2012nonadoption}. These studies are not autism-specific, but they provide a basis for analyzing \emph{monitored independence}: arrangements that relax physical proximity while preserving parental visibility and intervention power. They also make withdrawal consequential---who can cover, disable, refuse, or delimit monitoring, and under what conditions.

Privacy negotiation also extends beyond device settings. Chinese parents of autistic children describe strategic disclosure across home, school, and online contexts, balancing protection from stigma with access to understanding and support~\cite{wang2025chineseprivacy}. Such accounts illuminate parental practices but do not establish children's own experiences. This distinction is central to our evidence base: caregiver accounts can show how adults interpret and respond to children's actions, but they remain different from direct child perspectives. The same caution applies to China-based public platform materials, which can reveal situated family reasoning but cannot verify diagnoses or justify reproducing identifiable accounts.

\subsection{Configuration, Maintenance, and Care Work in the Smart Home}
\label{sec:rw-care-labor}

Domestic technologies depend on work that recedes from view when devices function routinely. \citet{tolmie2007digitalhousekeeping} describe the configuration, troubleshooting, and coordination required to maintain home networks as ``digital housekeeping.'' Smart-home studies show how this work can concentrate expertise and influence among ``pilot'' users, while ``passenger'' users live with arrangements they did not configure~\cite{koshy2021passenger}. Incomplete understandings of platforms, third-party services, and data flows further complicate informed configuration~\cite{abdi2019assistants}. These concepts recast system operation as an accomplishment distributed across devices and household members, rather than as the autonomous performance of a product.

In caregiving households, this work is also part of care. Protection can motivate smart-home adoption and redistribute domestic labor~\cite{strengers2019protection}, while household members describe configuration and maintenance through responsibility, affection, technical interest, and family support~\cite{gruhlich2025digitalcare}. Beyond installation and repair, domestic cognitive labor includes planning, coordination, anticipation, and judgment~\cite{frampton2026domestic}. For households of autistic children, these perspectives invite attention to additional forms of work: interpreting preferences, translating expressions into machine-recognizable commands, configuring sensory outputs, teaching device states, anticipating and repairing failures, and governing limits around time, content, intensity, and control. They also direct attention to unequal technical capacity and to who absorbs this work. Whether automation reduces, redistributes, or increases caregiver labor must therefore be established through ongoing household practice rather than inferred from adoption alone.

Prior work separately establishes possibilities for autism-related support, negotiations around family smart homes, and the often invisible labor of maintaining domestic technologies. Four intersections remain underexamined when families appropriate consumer devices around autistic children: how adaptation makes sensory environments, prompts, and interaction channels supportive; how general-purpose interaction rules and opaque states produce misfit and require recovery; how translation, configuration, maintenance, and boundary-setting become family care work; and how monitoring simultaneously enables movement and constrains privacy, agency, and withdrawal. These four analytic concerns organize our three research questions: RQ1 addresses adaptation and misfit, RQ2 addresses family work and its distribution, and RQ3 addresses monitored independence and household control. We examine them through caregiver interviews and autism-context Xiaohongshu accounts, using Weibo only as a general-family comparison. This source distinction allows us to analyze mechanisms and boundary conditions without treating public posts as clinical verification or caregiver interpretations as children's direct perspectives.

%% file: methodology.tex
\section{Methodology}
\label{sec:method}

\subsection{Research Design and Study Overview}

We used a multi-source qualitative research design to examine how smart-home technologies are experienced, used, adjusted, and limited in families of autistic children. The study integrated five caregiver interviews with publicly available social media posts from Xiaohongshu and Weibo. The three corpora were not pooled into a single dataset. Instead, each source was analyzed separately to preserve its interpretive logic, and the resulting source-level analyses were then compared in a cross-source joint analysis. The analysis remained attentive to the conditions under which support emerged, the work required to sustain it, and the tensions created by safety and privacy arrangements. Interviews served as the primary evidence for caregiver accounts of autistic children's everyday experiences and household decisions, and as core explanatory evidence for family processes and negotiation. Xiaohongshu contained both posts that explicitly situated experiences in autism-related contexts and posts that did not. The former served as naturalistic autism-related evidence; the latter were used only as adjacent or general-family contextual evidence. Weibo served solely as general-family context and comparison and was not used to support claims about autistic children. The hierarchy of the sources was determined by how directly each could address the research questions rather than by corpus size.

The analysis followed the phases of thematic analysis proposed by Braun and Clarke and was implemented as an inductive-deductive reflexive thematic analysis~\cite{braun2006using,braun2019reflecting}. The approach was inductive in that codes and themes were generated from and revised against the material. It was deductive in that sensory fit, communicative interaction, device state, family adaptation work, monitoring, safety, and privacy were used as initial analytic sensitivities. The analytic goal was not to estimate the prevalence of experiences but to identify the conditions under which smart-home technologies support or disrupt daily life and to trace the mechanisms through which families experience those conditions.

\begin{figure*}[t]
  \centering
  \includegraphics[width=\textwidth]{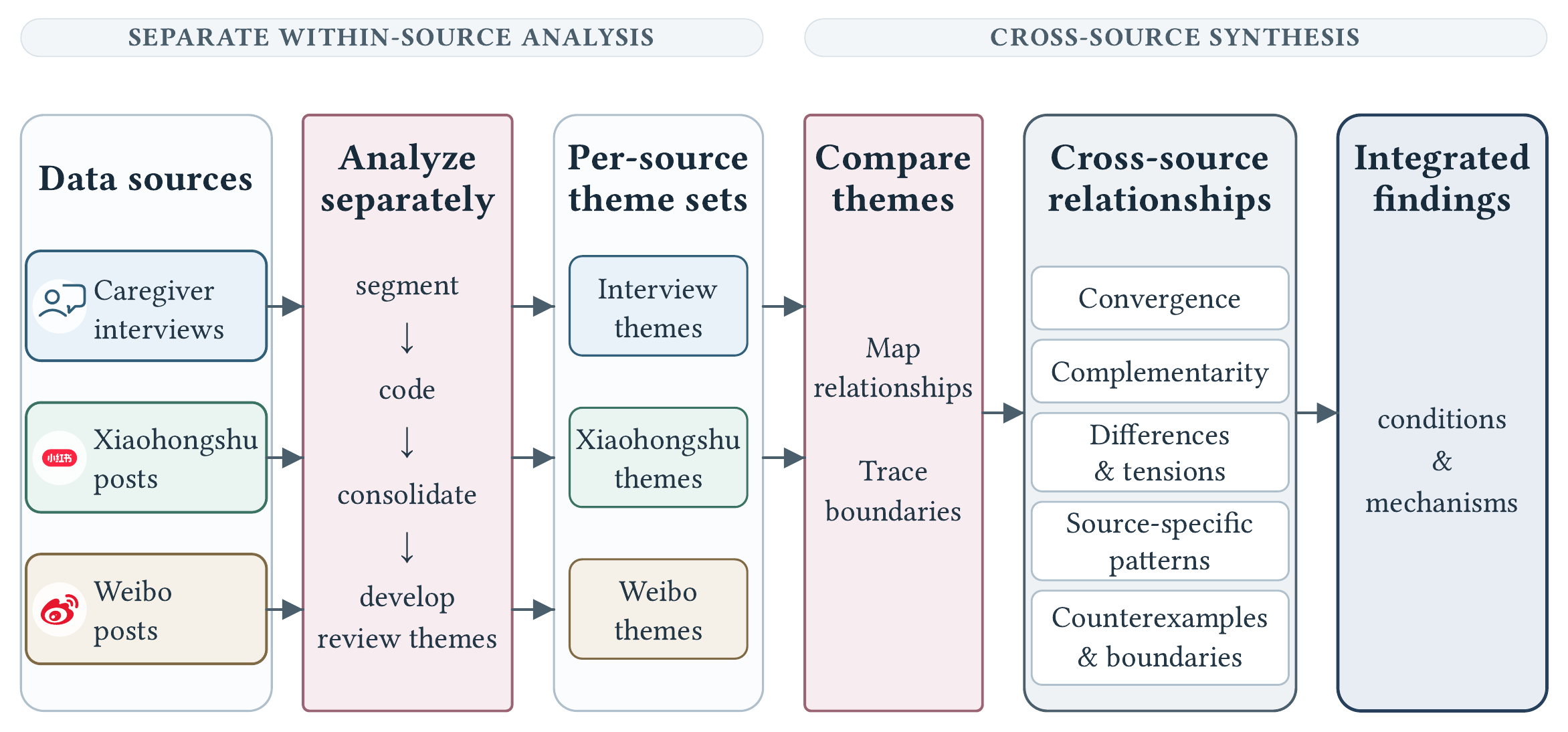}
  \caption{Multi-source study design and analytic synthesis.
  Caregiver interviews, Xiaohongshu posts, and Weibo posts were analyzed separately to produce source-specific theme sets. Interview and autism-context Xiaohongshu evidence formed the core cross-source comparison. General-family or adjacent Xiaohongshu evidence and Weibo evidence provided supplementary contextual checks. The comparison examined convergence, complementarity, differences and tensions, source-specific patterns, and counterexamples and boundaries, informing integrated findings about conditions and mechanisms.}
  \Description{A left-to-right flow diagram. Caregiver interviews, Xiaohongshu posts, and Weibo posts each point to an Analyze separately box containing the steps segment, code, consolidate, develop, and review themes. The analysis produces separate interview, Xiaohongshu, and Weibo theme sets. Interview and Xiaohongshu theme sets point to a Compare themes box containing map relationships and trace boundaries; the Weibo theme set remains visually separate to indicate its supplementary contextual role. The comparison leads to cross-source relationships comprising convergence, complementarity, differences and tensions, source-specific patterns, and counterexamples and boundaries, followed by integrated findings about conditions and mechanisms.}
  \label{fig:study-design}
\end{figure*}

Figure~\ref{fig:study-design} summarizes how the three data sources were analyzed separately and how their source-specific themes informed the cross-source synthesis.

\subsection{Participants and Interviews}

Five mothers of autistic children participated in the interview study. Participants were recruited through study notices posted on Xiaohongshu. To be eligible, a caregiver had to be the mother of an autistic child and to have experience with the child's or family's use of smart-home devices. Recruitment was not restricted to a particular device brand, household configuration, or child age beyond the study's focus on childhood smart-home use. The children represented in the sample were between 8 and 11 years old. All participants were female caregivers and were mothers of the focal child.

Data collection took place between 24 June and 15 July 2026. Interviews were conducted in Chinese using a semi-structured interview guide. The guide covered the family's device ecology, the child's sensory and communicative responses to devices, caregiver adaptation and maintenance practices, monitoring and safety arrangements, and experiences of privacy and control. The interviewer followed the guide but used flexible follow-up probes to clarify concrete incidents, sequences of action, and the caregiver's interpretation of the child's response.

Two Chinese-speaking students in the human--computer interaction (HCI) field conducted the interviews. One conducted the three interviews in the first recruitment wave, and the second conducted the two interviews in the second wave. Contact with participants began through the recruitment process, and no participant had a pre-existing relationship with either interviewer. Two interviews were conducted by voice; they lasted approximately 30 and 43 minutes, respectively, and were audio-recorded with consent. The recordings were transcribed using a text transcription application, and the resulting Chinese transcripts served as the analytic record. The other three interviews were conducted through synchronous text-based exchanges: one was a WeChat text interview, and two were text-based questionnaire interviews. Three of the text-based interviews lasted approximately 30 minutes each. The text-based exchanges did not extend across multiple days. No monetary or other compensation was provided. Interview records were anonymized as P1 to P5 before analysis, with identifying institutional, geographical, and family details removed.

The study planned five caregiver interviews, and recruitment was closed once all five had been completed. Five cases were sufficient for the study's exploratory, mechanism-oriented aims because each caregiver contributed a distinct family context, device ecology, and account of how support was produced or disrupted; the analysis used cross-case comparison and counterexamples rather than frequency estimation. This sample size does not support claims about prevalence or generalization to all autistic children or families, and we do not claim empirical saturation. 

Table~\ref{tab:participants} summarizes the participant roles, child age range, principal devices, interview modes, and interview extent. Exact child ages are reported at the sample level rather than assigned to individual cases to reduce re-identification risk.

Each participant was treated as a case whose family circumstances, device ecology, and reported practices could differ from those of the other participants. The caregiver in P1 described a comparatively rich history of device modification and integration; we report that case as an information-rich case rather than as evidence of common family practice across the sample. Segmentation of the interview material produced 381 initial meaning units. After student researchers' review, code consolidation, and removal of technical duplicates, 348 units were retained as thematic evidence, supported by 38 focused codes that emerged from an initial set of 89 codes. The interview analysis yielded five within-source themes.

One participant described a child as having Asperger syndrome based on the caregiver's account of physical examinations and observed characteristics. Because this description was not formally diagnosed by a clinician, we treat it as caregiver-identified and do not generalize from it as a clinical category.

\begin{table*}[t]
  \caption{Participant and interview characteristics.}
  \label{tab:participants}
  \centering
  \small
  \begin{tabular}{@{}p{0.05\textwidth}p{0.10\textwidth}p{0.11\textwidth}p{0.27\textwidth}p{0.17\textwidth}p{0.16\textwidth}@{}}
    \hline
    ID & Caregiver role & Child age range & Main devices & Interview mode & Duration/extent \\
    \hline
    P1 & Mother & 8--11$^{*}$ & Smart lighting, speaker, connected sensors, robot vacuum, smart lock, and location wearable & Voice (recorded) & 43 min \\
    P2 & Mother & 8--11$^{*}$ & Smart speaker (Xiaodu) and smart water heater & Voice (recorded) & 30 min \\
    P3 & Mother & 8--11$^{*}$ & Smart lighting, speaker/AI assistant, smartwatch, smart lock, and cameras & Text (synchronous) & 180 min \\
    P4 & Mother & 8--11$^{*}$ & Smartwatch; smart speaker considered but not adopted; household cameras & Text (synchronous) & 30 min \\
    P5 & Mother & 8--11$^{*}$ & Smart speaker, cameras, smart lock, connected washing machine, and physical controls & Text (synchronous) & 30 min \\
    \hline
  \end{tabular}

  \par\vspace{2pt}
  \begin{minipage}{0.95\textwidth}
  \footnotesize $^{*}$The age range is reported for the sample as a whole; individual ages were not retained in the reporting table to reduce re-identification risk. 
  \end{minipage}
\end{table*}

\subsection{Social Media Data Collection and Screening}

Social media records were collected between 10 and 19 August 2026, using Octoparse to perform keyword-based scraping of publicly accessible posts on Xiaohongshu and Weibo. We used each platform's native keyword search rather than random sampling. The queries combined Chinese terms for autism or autistic family contexts (for example, \emph{zibizheng}, \emph{guduzheng}, \emph{zibizheng ertong}, and \emph{zibizheng jiating}) with smart-home and device terms (for example, \emph{zhineng jiaju}, \emph{zhineng yinxiang}, \emph{zhineng deng}, \emph{zhineng mensuo}, \emph{zhineng shoubiao}, \emph{shexiangtou}, and \emph{saodi jiqiren}). For the second Weibo collection, child- and family-related terms (for example, \emph{haizi}, \emph{ertong}, and \emph{jiazhang}) were combined with the same device vocabulary. These are representative Chinese search terms; the complete query strings are retained in the project audit log. Because platform search results are relevance-ranked and can vary with account, time, and personalization, the corpus is a bounded sample of search-accessible public posts rather than an exhaustive census of either platform.

Within each returned record, the focal post was the unit of retrieval and screening; comments and replies were not included in the corpus. Posts had to be publicly accessible and had to document an actual personal or family experience involving smart-home use and children; advertising, marketing, news, pure popular-science content, institutional promotion, low-information posts, and unrelated material were excluded. Duplicate records were defined as records sharing the same platform post identifier or canonical URL. Near-identical reposts of the same content were also collapsed during the second screening pass. Multiple eligible posts by the same author were retained as separate content records when they described distinct experiences or device interactions, but they were linked through an author-level identifier so that repeated accounts from one account were not treated as independent participants. Posts that had become deleted, restricted, or inaccessible before screening were excluded because their content could not be verified. No attempt was made to access private accounts or restricted content; posts that were public at collection but later deleted remained only in the de-identified internal corpus and are not reproduced through usernames, profile information, or links.

Screening was conducted by two native Chinese-speaking researchers. One completed the initial screening pass, and the second independently checked all inclusion and exclusion decisions; disagreements were resolved by discussion against the eligibility criteria. The Xiaohongshu corpus was retrieved through autism-related and smart-home-related searches. Not every retained post explicitly identified the child or family as autistic, so the platform corpus was not treated as uniformly autism-specific. At the evidence-use stage, posts that explicitly situated the experience in an autism context were used as autism-related naturalistic evidence; posts without that context were used only as adjacent or general-family contextual evidence. After removal of exact duplicate records, the initial collection contained 316 unique posts. A first screening pass classified posts into personal experience, product promotion, popular science, news, institutional promotion, academic or industry information, questions or opinion discussion, and unrelated content. The two screeners retained 116 posts as personal-experience candidates. A second pass removed residual promotional material, low-information content, posts without direct relevance to children and smart-home use, and near-duplicate reposts, leaving 76 eligible personal-experience posts. Of these, 47 explicitly situated the reported experience in an autism-related context and were used as autism-related evidence; the remaining 29 were used only as adjacent or general-family context. Segmentation produced 330 meaning units, which were reduced to 287 after code consolidation and technical deduplication. The analysis produced seven within-source themes.

We initially searched Weibo using combinations of autism-related terms and smart-home or device-related terms. This search returned 424 records, but most were technology news, public-interest or promotional materials, industry information, figurative uses of autism-related language, or posts unrelated to smart-home use. Only five records contained relevant personal experience, and these provided insufficient detail for coding; we therefore did not conduct thematic analysis on this subset. We subsequently broadened the search to combinations of child-related and smart-home-related terms. This search returned 1,140 records, and 245 eligible personal-experience posts remained after screening by the two native Chinese-speaking researchers. Posts were classified into the same content categories as the Xiaohongshu corpus, and only personal or family interaction experiences were retained. The final coded workbook contained 443 coding records, which were consolidated into 382 deduplicated meaning units and 87 focused codes, organized into six within-source themes.

The two platforms are therefore not symmetric evidence. Xiaohongshu contributes naturalistic autism-related accounts when the post text makes that relevance explicit; without such relevance, it contributes only general-family or adjacent technology material. Weibo contributes general family context and boundary conditions. Weibo posts cannot be used to infer that a child is autistic, and neither platform's tags nor authors' self-descriptions were treated as clinical verification.

Table~\ref{tab:study-materials} summarizes the participant and material contexts, retained material, and evidentiary roles of the three sources.

\begin{table*}[t]
  \caption{Study materials, participant context, and evidentiary roles.}
  \label{tab:study-materials}
  \centering
  \input{figures/T1_study-materials_body_v1.1.tex}
\end{table*}

\subsection{Data Analysis}

The analytic procedure had seven stages. First, the student researchers read the complete Chinese interview transcripts and social media posts repeatedly. For each interview, a case-level memo recorded family background, actual device use, child differences, caregiver judgments, and contradictions.

Second, the analytic corpus was screened so that only material directly related to smart-home use and child-device interaction was retained. Relevant material covered adaptation and misfit; translation, configuration, modification, maintenance, and boundary-setting; monitoring, safety, and privacy; support and difficulty; and concrete design needs. Advertising, marketing, news, pure popular science, institutional promotion, low-information expression, and unrelated content were excluded.

Third, interview and post content was divided into semantically complete meaning units. A single post or interview segment could produce multiple meaning units, but each unit retained its full expression, participant or post identifier, source location, and enough context to be understood independently. 

Fourth, the student researchers generated short, data-near initial codes. Sensory fit, communicative interaction, device state, family adaptation work, monitoring, safety, and privacy were used as initial sensitizing concepts, while the analysis remained open to codes outside these categories. Actual experiences, evaluative statements, unmet needs, and future expectations were tagged separately so that hopes for technology were not reported as demonstrated effects.

Fifth, semantically close initial codes were consolidated into focused codes and organized under three broad analytic areas: adaptation and misfit; family work; and monitoring, safety, and privacy. Analytic memos recorded the phenomenon represented by each code, its possible mechanism, its boundary conditions, its relation to other codes, cross-case variation, and counterexamples. Only technical duplicates created by code consolidation were removed; similar experiences from different families or posts were retained because their recurrence was the basis for identifying patterns. 

Sixth, candidate themes were generated and reviewed within each source. Themes were assessed for internal coherence, distinction from other themes, breadth across participants or posts, relevant counterexamples, and fit with the research questions. Counts were used only as retrieval and distribution tools and were not interpreted as prevalence, importance, or saturation.

Seventh, the three source-level analyses were compared directly. The comparison examined convergence, in which different sources supported the same experience; complementarity, in which different sources explained different layers of the same phenomenon; differences and tensions; source-specific findings; and counterexamples and boundary conditions. These comparisons produced the four cross-source themes reported in the Findings; they were not treated as source themes placed side by side. Interview and autism-context Xiaohongshu evidence supported the development of the themes, while general-family or adjacent Xiaohongshu evidence and Weibo evidence were used to examine their scope and boundary conditions. The Discussion subsequently interprets these themes through three related concepts: relational fit, family work as the infrastructure of support, and monitored independence.

For the general-family Weibo corpus, we additionally assessed whether experiential elements such as light, noise, unexpected device output, controllability, and safety were analytically relevant to questions raised in the autism-focused materials. This assessment concerned the relevance of the reported interaction or environmental feature, not the diagnostic identity of the poster or child. We did not infer autism status from these posts or use them as autism-specific evidence.

\subsection{Roles of the Three Data Sources and Claim Boundaries}

The data sources have distinct evidentiary roles. Interviews are the core explanatory source because they allow the analysis to trace why families adopt, adjust, or reject technology, how decisions are made, and how safety and privacy boundaries are negotiated. Xiaohongshu provides naturalistic autism-related evidence when a post explicitly situates the experience in that context; posts without such context provide only adjacent or general-family material. Weibo is a supplementary context and comparison source because it shows how smart-home interactions, maintenance practices, and device failures occur in ordinary child-raising households.

This division of roles constrains the claims that can be made. Weibo alone cannot support conclusions about autism. P1's detailed practices cannot be written as shared practice across all five families. The caregiver-identified account of Asperger syndrome cannot be treated as a clinically confirmed diagnosis. Counts of meaning units and codes describe the depth and distribution of the analyzed material, not the frequency of an experience in the broader population. We therefore restrict claims to patterns, conditions, and mechanisms present in the material rather than generalizing to all autistic children, all families, or all smart-home technologies.

\subsection{Ethical Considerations}

The interview component was not subject to local institutional review board (IRB) review requirements because no local IRB or equivalent review process existed at the time of data collection.

The interview component involved five caregivers; children were not directly interviewed. Participants provided informed consent through electronic communication, and records documenting that consent were retained.

Interview materials used in analysis and reporting were labeled P1 to P5. Identifying institutional, geographical, and family details were removed or generalized during preparation for reporting, and interview quotations are attributed only to participant-level identifiers. Because the sample is small and the interviews concern potentially sensitive information about children and family life, we avoid reporting combinations of contextual details that could identify a family.

The social media corpus comprised Xiaohongshu and Weibo posts that were publicly accessible at the time of collection. We treated public accessibility as an access condition rather than as evidence that authors expected their posts to be reproduced in research. Source links were retained in internal records for traceability, but the paper does not report usernames, profile information, links, or full post texts. Social media excerpts are presented as English translations of the Chinese originals, not as paraphrases. These translated quotations were checked against the source text and are attributed using study-assigned identifiers. Translation and the omission of direct identifiers do not eliminate the risk of re-identification from distinctive quoted content, especially where posts describe children or autism-related family life. We did not treat platform tags, authors' self-descriptions, or caregiver accounts as clinical verification. Xiaohongshu material is described as autism-related only when the post text made that relevance explicit; Weibo is used only as general child-raising context and is not used to infer that a child is autistic.

The student researchers read the Chinese source material directly and approached the data with explicit attention to sensory regulation, communication accommodation, care work, child agency, and privacy. These commitments were treated as interpretive resources requiring discipline rather than as neutral categories. To support auditability, each retained analytic unit preserves its source identifier, original location, code history, and analytic memo reference, and the final coding structure is documented separately from the paper.

The first author was responsible for initial screening, meaning-unit segmentation, initial coding suggestions, code consolidation, evidence retrieval, and drafting.
 
Chinese originals were treated as authoritative, and English quotations were checked against them. 
Manual work included re-segmenting incomplete meaning units, correcting codes, removing advertising and irrelevant content, merging technical duplicates, verifying representative evidence, and making the final decisions about inclusion, coding, and interpretation.

%% file: figures/T1_study-materials_body_v1.1.tex
\begingroup
\normalsize
\setlength{\tabcolsep}{4pt}
\renewcommand{\arraystretch}{1.16}
\definecolor{TOneHeaderColor}{HTML}{EEF2F5}
\begin{tabularx}{\linewidth}{@{}
  >{\raggedright\arraybackslash}p{0.12\linewidth}
  >{\raggedright\arraybackslash}p{0.21\linewidth}
  >{\raggedright\arraybackslash}p{0.12\linewidth}
  >{\hsize=1.08\hsize\linewidth=\hsize\raggedright\arraybackslash}X
  >{\hsize=0.92\hsize\linewidth=\hsize\raggedright\arraybackslash}X@{}}
  \toprule
  \rowcolor{TOneHeaderColor}
  \textbf{Source} & \textbf{Participant / material context} & \textbf{Included material} & \textbf{Evidentiary role} & \textbf{Key boundary} \\
  \midrule
  \textbf{\mbox{Caregiver}\newline\mbox{interviews}}
  & Female caregivers of autistic children aged 8--11
  & 5 caregivers
  & \textbf{Core explanatory}\newline Family decisions, technology adjustments, and negotiation
  & Caregiver accounts; children were not directly interviewed \\
  \addlinespace[7pt]
  \textbf{Rednote}
  & Retained personal-experience posts from autism- and smart-home-related searches
  & 76 posts
  & \textbf{Core naturalistic}\newline When the ASD context was explicit; otherwise adjacent or general-family context
  & Not all posts explicitly identified an ASD context; diagnoses were not clinically verified \\
  \addlinespace[7pt]
  \textbf{Weibo}
  & Personal or family-interaction posts in general child-raising contexts
  & 245 posts
  & \textbf{Supplementary comparison}\newline Family interactions, maintenance, and device failures
  & Cannot independently support conclusions about autism \\
  \bottomrule
\end{tabularx}
\par\smallskip
{\footnotesize\raggedright
\textit{Notes.} Interviews comprised two voice interviews and three text-based written exchanges.  Xiaohongshu refers to Rednote. Counts refer to caregivers or retained posts; posts do not necessarily represent distinct people or households.\par}
\endgroup